\documentclass[12pt]{article}
\usepackage{epsfig}
\begin{document}
\begin{titlepage}
\begin{center}
\vspace{2cm}
 {\Large \bf
Instanton-Induced Polarization \\
in Exclusive Hyperon Photoproduction}
\vspace{0.50cm}\\

Nikolai Kochelev$^{a,b}$\footnote{kochelev@theor.jinr.ru} \vspace{0.50cm}\\
{(a) \it School of Physics and Astronomy, Seoul National University,\\ Seoul 151-747,  Korea}\\
{(b) \it Bogoliubov Laboratory of Theoretical Physics, Joint
Institute for Nuclear Research, Dubna, Moscow region, 141980
Russia} \vskip 1ex
\end{center}
 \vskip 0.5cm \centerline{\bf Abstract} It is shown that
 instantons can provide an explanation of the strong polarization
 transfer observed in exclusive hyperon photoproduction at JLab.
\vspace{1cm}

\end{titlepage}

\setcounter{footnote}{0} In recent exclusive hyperon
photoproduction experiments at JLab a remarkably strong transfer
of polarization from the initial photon to the final hyperon has
been observed in the reactions $\gamma+p\rightarrow K^++\Lambda$
and $\gamma+p\rightarrow K^++\Sigma^0$
\cite{Bradford:2006ba,Ambrozewicz:2006zj}. The QCD mechanism
underlying this polarization transfer remains unexplained; to
date, only a qualitative, phenomenological explanation has
appeared in the literature \cite{Schumacher:2006ii}, which is
based on an {\it ad hoc} assumption of complete spin transfer from
the photon to the strange quark. In this Brief Report we discuss a
possible instanton-based mechanism for the strong correlation
observed between the helicities of the initial photon and final
hyperon.

Instantons, strong vacuum gluon fields which describe sub-barrier
transitions between topologically different vacua in QCD, are
known to play an important r\^ole in the dynamics of the strong
interaction (for a review see \cite{Schafer:1996wv}). In
particular, instantons induce the 't~Hooft four-quark interaction,
which for massless quarks is given by \cite{Shifman:1979nz}
\begin{eqnarray}
{\cal H}_{^{\prime}t\, Hooft}& =&
\int\! d\rho \, n(\rho)\, (\frac{4}{3}\pi^2\rho^3)^2 \,
\{\bar u_{R}u_{L}\bar s_{R}s_{L}
[1+\frac{3}{32}(1-\frac{3}{4}
\sigma^u_{\mu\nu}\sigma^s_{\mu\nu})\cdot \lambda^a_u\lambda^a_s]\\
\nonumber &+&(u\to d)+(s\to d)+(R\to L) \},
\label{thooft}
\end{eqnarray}
where $n(\rho)$ is the instanton density and $\rho$ is the instanton
size. Here we shall assume that the instanton-induced interaction (1)
is responsible for the observed strong correlation between the
chiralities of the initial and final quarks.

The interaction (1) specifies that the initial quarks interacting
through the instanton (anti-instanton) have left-handed
(right-handed) chiralities, and that the final quarks have the
opposite right-handed (left-hand) chiralities. For massless quarks
chirality coincides with helicity, so we anticipate fully
correlated initial and final helicities for the quarks interacting
through the instanton field.

\begin{figure}[htb]
\centering
 \psfig{file=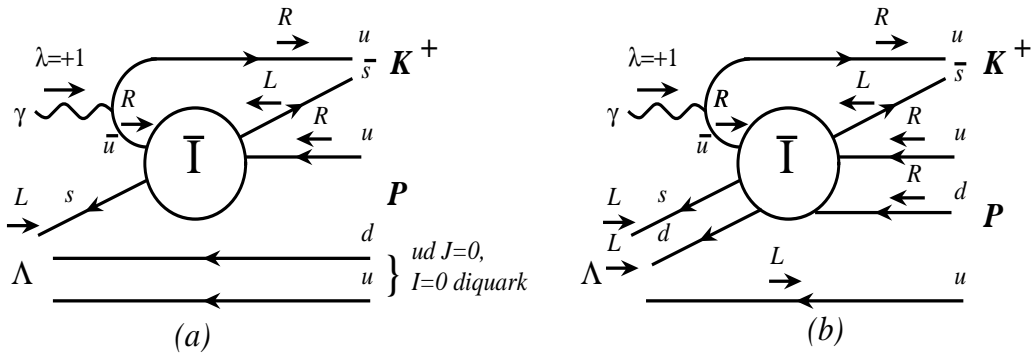,width=14cm, height=6cm} \caption{ The
instanton mechanism for polarization transfer to a $\Lambda$
hyperon: a) four-quark and b) six-quark instanton induced
interaction contributions. Here I denotes the instanton, R and L
are the quark helicities, and the detached arrows show the
directions of the quark spins.}
\end{figure}

For definiteness we consider exclusive instanton-induced $\Lambda$
photoproduction due to interaction (1) in the center of mass
system (Fig.1a). We emphasize that the same coordinate system was
been used  by the JLab experiment
\cite{Bradford:2006ba,Ambrozewicz:2006zj} for analysis of the
photon to $\Lambda$ helicity transfer. Inspection of the figure
shows that the helicities of the initial light quarks which couple
to the photon will be opposite to the helicities of the final
$s\bar s$ quark pair. Since the polarization of the
$\Lambda$-hyperon is given by the polarization of the $s$ quark
(in the valence approximation), the $\Lambda$ should be fully
polarized in the direction of the photon helicity.\footnote{The
case of $\Sigma^0$ photoproduction is more complicated because the
spin state of this hyperon is not determined by the strange quark
alone.} We note in passing that there is an additional instanton
contribution to this reaction due to the six-quark
instanton-induced interaction (Fig.1b). Although this additional
contribution leads to the same strange quark helicity orientation
relative to the photon chirality as the four-quark interaction
(1), it should be strongly suppressed in scale. This six-quark
interaction requires a double quark spin flip (of the d- and
s-quarks in Fig.1b); one cannot combine the final quarks to form a
spin-1/2 hyperon in the non-relativistic limit. It is also clear
that the one-instanton mechanism may only be applicable at rather
low value of energy excess $Q$ above threshold.
($Q=W-M_{\Lambda}-M_K\leq M_S$, where $M_S\approx 2~$GeV is the
so-called sphaleron energy, which is the height of the potential
barrier between different vacua \cite{Janik:2002nk}. At higher
energies, multiinstanton contributions should lead to a
decorrelation of hyperon and photon helicities. Due to the
suppression of multiinstanton contributions at large photon
virtuality, one might anticipate that the one-instanton
approximation is better justified in exclusive electroproduction,
in which a strong correlation between photon and hyperon helicity
has also been observed \cite{Carman:2002se}. In this case however
one should carefully separate the zero-helicity longitudinal
photon contribution.

In our conjecture the valence quark approximation for hadron and
photon wave functions is using. It is evident that sea quark
contributions should reduce helicity correlations at high energy.
In the context of the instanton model of the QCD vacuum, these
non-valence components in hadron state vectors are related to the
admixture of quark-antiquark pairs and gluons created by
instantons. This effect is not incorporated in our
single-instanton model of hyperon production at energies not far
above threshold.

The validity of this instanton mechanism for polarization transfer
can be tested through studies of exclusive hyperon photoproduction
in association with a vector meson ($K^*$). In this case we
anticipate fully longitudinal polarization of the $K^*$, due to
the quark chirality flip at the instanton vertex (see Fig.1).

In summary, we have suggested a new mechanism for polarization transfer
in exclusive hyperon photoproduction, based on 't~Hooft's instanton-induced
four-quark interaction. This mechanism can be tested through studies of
hyperon photoproduction associated with a strange vector ($K^*$) meson.

\vskip 1cm

We are happy to acknowledge useful discussions of this work with
A.E.Dorokhov. The author is very grateful to the School of Physics
and Astronomy of Seoul National University, and especially Prof.
Dong-Pil Min, for their kind hospitality. This work was supported
in part by the Brain Pool program of the Korea Research Foundation
through KOFST grant 042T--1--1.

\end{document}